\documentclass[conference]{IEEEtran}
\IEEEoverridecommandlockouts
\usepackage{cite}
\usepackage{amsmath,amssymb,amsfonts}
\usepackage{algorithmic}
\usepackage{graphicx}
\usepackage{textcomp}
\usepackage{xcolor}
\def\BibTeX{{\rm B\kern-.05em{\sc i\kern-.025em b}\kern-.08em
		T\kern-.1667em\lower.7ex\hbox{E}\kern-.125emX}}
\usepackage{tikz}
\usepackage{pgfplots}
\usepackage{pgfplotstable}
\pgfplotsset{compat=1.7}
\usepackage{float}
\usepackage{multirow}
\usepackage{subcaption}
\usepackage[]{algorithm2e}
\usepackage{blindtext}
\usepackage[framemethod=tikz]{mdframed}
\usepackage{lipsum}
\usepackage[normalem]{ulem}
\usetikzlibrary{fit} % Load the fit library for grouping nodes
\usetikzlibrary{positioning}
\colorlet{shadecolor}{blue!20}

\begin{document}
	
	\title{On Splitting Lightweight Semantic Image Segmentation for Wireless Communications}
	
	\author{
		\IEEEauthorblockN{
			Ebrahim Abu-Helalah\IEEEauthorrefmark{1}\IEEEauthorrefmark{2}, 
			Jordi Serra\IEEEauthorrefmark{1},
			Jordi Perez-Romero\IEEEauthorrefmark{2}, 
		}
		\IEEEauthorblockA{\IEEEauthorrefmark{1}Centre Tecnol\`{o}gic de Telecomunicacions de Catalunya (CTTC/CERCA), Barcelona, Spain}
		\IEEEauthorblockA{\IEEEauthorrefmark{2}Universitat Politecnica de Catalunya (UPC), Barcelona, Spain\\
			\{aebrahim, jserra\}@cttc.es, \{ebrahim.abu.helalah,jordi.perez-romero\}@upc.edu}
	}
	
	\maketitle
	
	\begin{abstract}
		Semantic communication represents a promising technique towards reducing communication costs, especially when dealing with image segmentation, but it still lacks a balance between computational efficiency and bandwidth requirements while maintaining high image segmentation accuracy, particularly in resource-limited environments and changing channel conditions. On the other hand, the more complex and larger semantic image segmentation models become, the more stressed the devices are when processing data. This paper proposes a novel approach to implementing semantic communication based on splitting the semantic image segmentation process between  a resource constrained transmitter and the receiver. This allows saving bandwidth by reducing the transmitted data while maintaining the accuracy of the semantic image segmentation. Additionally, it reduces the computational requirements at the resource constrained transmitter compared to doing all the semantic image segmentation in the transmitter. The proposed approach is evaluated by means of simulation-based experiments in terms of different metrics such as computational resource usage, required bit rate and segmentation accuracy. The results when comparing the proposal with the full semantic image segmentation in the transmitter show that up to 72\% of the bit rate was reduced in the transmission process. In addition, the computational load of the transmitter is reduced by more than 19\%. This reflects the interest of this technique for its application in communication systems, particularly in the upcoming 6G systems.
	\end{abstract}
	
	\begin{IEEEkeywords}
		semantic communication; semantic image segmentation; artificial intelligence; machine learning; 6G. 
	\end{IEEEkeywords}
	
	\section{Introduction}
	
	The Ericsson Mobility Report from November 2024\cite{ref01-ericsson2024mobility} predicts that by 2030, there will be around 6.3 billion 5G mobile users worldwide. Moreover, there are substantial efforts in starting to shape the future 6G networks that will target challenging usage scenarios such as hyper-reliable and low-latency communication, immersive communication or ubiquitous connectivity\cite{ref02-iturm21600}. It will also focus on artificial intelligence assistance in communications, encompassing standard applications such as automated driving, autonomous device collaboration for medical assistance, and distributing intensive computing tasks across devices and networks\cite{ref02-iturm21600}. In light of this expected huge increase in the number of users and traffic demand and preparation for 6G network, it is necessary to develop communication models capable of transferring data faster and more efficiently.
	
	\vspace{1pt}
	
	Even though traditional communications have come a long way, they still have a hard time keeping up with the exponential growth in data traffic and the growing need to send large amounts of data, i.e., the bandwidth bottleneck. These systems transmit large amounts of raw data, which strains the channel capacity and leads to problems such as network congestion and increased latency\cite{ref03-goldsmith2005wireless}. This is a particular problem in real-time applications, such as live video streaming or autonomous systems, where even small delays can obstruct performance\cite{ref04-servedcosyto2019}. For instance, streaming a 4K video may require 30-50 Mbps\cite{ref13-uhd4k2021}, and virtual reality applications may require 25-100 Mbps for each user\cite{ref14-mangiante2017VR}. In this context, providing service to a large number of users leads to unprecedented demands in network infrastructure. Hence the need for a smarter approach that prioritizes the meaning of data over raw data. In this direction, semantic communication emerges as an effective model in meeting this objective\cite{ref05-semcom62562202410693810iccc}.
	
	\vspace{1pt}
	
	In contrast to traditional communication, which focuses on transferring raw data without knowing their meaning or significance, i.e., pre-understanding transfer, semantic communication emphasizes the accurate transfer of meaning behind data, known as post-understanding transfer. The principles of semantic communication are widely applicable to systems that require intelligent and meaningful exchange of data\cite{ref06-semcom1012100269openissue}.
	
	\vspace{1pt}
	
	Semantic communication addresses the issue of bandwidth limitations by concentrating on the transmission of only the essential semantic information, thereby lowering the necessary bit rate. However, this approach presents a new challenge, which is the limitation in computational resources. Transforming raw data into valuable insights, such as semantic segmentation, demands significant computing power \cite{ref07-qin2022semanticcommunicationsprincipleschallenges}. Regardless of using lightweight models and optimization methods, the computational demands are still significant, particularly in environments with limited resources like when the transmitter is a mobile terminal \cite{ref07-qin2022semanticcommunicationsprincipleschallenges}. The challenge of maintaining a balance between limited bandwidth and growing computational needs presents a significant obstacle to the broader implementation of semantic communication.
	
	\vspace{1pt}
	
	Among the applicability areas of semantic communication, semantic image segmentation involves creating a detailed map of an image, labeling each part with its corresponding meaning. Instead of sending the entire image, we can just send the labels for those meaningful parts, for example, “sky,” “tree,” “car,” “person” \cite{ref06-semcom1012100269openissue}. This approach allows us to focus on the key information within the image, making communication more efficient and easier to understand. Imagine describing a scene to someone instead of showing them a blurry image—semantic segmentation works similarly by providing a concise, meaningful representation of the visual world \cite{ref08-fullconvsemcomnet2015}.
	
	\vspace{1pt}
	
	Semantic communication and semantic image segmentation, including the lightweight models, are important aspects in the development of intelligent systems based on post-understanding transmission. Despite the supposed interconnection between these two areas, previous research has often addressed them separately, leading to a research gap that prevents the desired integration. For example, papers such as Proportional-Integral-Derivative Network (PIDNet) \cite{ref09-pidnet01} and Lightweight Context-Aware Network (LCNet) \cite{ref10-lcnet01} have focused on developing lightweight semantic segmentation without paying attention to their integration with semantic transmission frameworks. Other papers, like \cite{ref11-e2e2023}, which looks at semantic transfer using polar codes for semantic-based communication systems, or \cite{ref12-we2e2023}, which looks at wireless semantic communication for image transmission, have focused on the framework of semantic transmission without considering lightweight semantic encoder methods. This research gap, in addition to the problems mentioned above, highlights the need for a study that combines both aspects to achieve a better integration between lightweight semantic image segmentation and semantic transfer frameworks.
	
	\vspace{1pt}
	
	Based on the above, the aim of this paper is to investigate the application of lightweight image segmentation models within semantic communication pipelines. Specifically, we will explore the trade-offs between computational resources at the transmitter, communication resources, and application accuracy. To achieve this, the main contribution of this paper is the proposal of splitting a lightweight semantic encoder (distributing its components between the sender and the receiver) and analyzing its impact on both communication and computational resource consumption.
	
	\vspace{1pt}
	
	Specifically, the novelty of the paper is the application of PIDNet in semantic communication by splitting the network after the fifth stage. This approach allows the sender to transmit low-resolution, semantically rich features, while the receiver completes the semantic inference process using the final layers of the model. Unlike traditional partitioning or edge inference methods, our design explicitly considers the balance between feature size, which reduces the communication resources used, and semantic accuracy across the communication channel.
	
	\vspace{1pt}
	
	The rest of the paper is organized as follows: Section \ref{probstate} presents a comprehensive description of the considered problem. Section \ref{probsul} presents the proposed solution that is experimentally evaluated in Section \ref*{expres}. Finlay, Section \ref*{conc} concludes this work.
	
	\vspace{1pt}
	
	\section{Problem Statement}\label{probstate}
	
	The problem addressed by this paper is to achieve the balance between accuracy, bit rate, and computational requirements of semantic image segmentation within the framework of semantic communication. Semantic image segmentation is a fundamental task in the field of computer vision that involves classifying each pixel in an image into a predefined class. Unlike image classification, which assigns a single label to the entire image, semantic segmentation provides a detailed understanding of a scene by distinguishing objects and their boundaries at the pixel level. To ensure high accuracy in image semantic segmentation, convolutional neural networks (CNNs) have been widely used. Where CNNs extract hierarchical spatial features from input images through layers of convolutions, nonlinear activations, and pooling operations \cite{ref08-fullconvsemcomnet2015}. However, these techniques also introduce different challenges, the most notable of which is dealing with objects of different sizes within the image, because CNN networks have a fixed receptive field, which limits their ability to capture multi-scale contextual information. The Pyramid Scene Analysis Network (PSPNet) introduces the Pyramid Pooling Module (PPM) to address this challenge \cite{ref18-pspnet2017}. The purpose of PPM is to combine contextual data at different levels by using global pooling. This allows to make several feature maps with varying levels of accuracy. This technique improves segmentation performance, especially in complex scenes containing objects of different sizes. Even though PPM has benefits, it comes with extra computing costs. This is because multi-scale pooling operations need more memory and processing power, which can be expensive for real-time applications, especially when working with high-resolution images.
	
	\vspace{1pt}
	
	Moreover, implementing this technology in a resource-limited environment poses significant challenges due to the computational requirements needed for extracting this meaning, e.g. through semantic segmentation. Devices like edge sensors, IoT nodes, and mobile platforms face strict limitations, such as low-performance processors, limited memory, and restricted power supply \cite{ref15-will6g2024}. For instance, a smart home sensor designed for real-time anomaly detection needs to process data locally while conserving battery life, which makes it difficult to use resource-heavy semantic algorithms. Similarly, drones that rely on image semantic segmentation for navigation must find a balance between lightweight processing models and the accuracy required to navigate. These examples underscore the difficulties of applying semantic communication systems in environments with limited computational resources.
	
	\vspace{1pt}

	The requirement to find an optimal compromise between the computational load and the complexity of CNNs has led to the design of new methods of semantic image segmentation. So, researchers have been developing lightweight neural networks (e.g., MobileNet \cite{ref16-mobilenet2017}, Tiny YOLO \cite{ref17-tinyyolo2023}, and PIDNet) to perform the AI tasks within resource-constrained devices with minimum processing cost, which could be implemented on edge devices. But these strategies may still suffer in the resource-limited environment. 
	
	\vspace{1pt}

	To achieve the balance mentioned earlier, we have proposed an innovative solution based on semantic encoder splitting between the transmitter and the receiver which will be discussed in the next section.
	
	\vspace{1pt}

	\section{A Splitting-Based Approach To Semantic Communication}\label{probsul} 
	
\begin{figure}[htbp]
		\centering
		\tiny
		\begin{tikzpicture}[scale=0.8, node distance=30mm and 0.9cm, align=center]
			% Transmitter Section
			\node [draw, minimum width=7mm, minimum height=0.8cm] (input) {\tiny$Input$\\\tiny$Image$};
			\node [draw, minimum width=7mm, minimum height=0.8cm, right=3mm of input] (semantic1) {\tiny$Semantic$\\\tiny$Encoder$\\\tiny$P1$};
			\node [draw, minimum width=7mm, minimum height=0.8cm, right=3mm of semantic1] (encoder) {\tiny$Radio$\\\tiny$Protocol$\\\tiny$Stack$};
			% Physical Channel
			\node [draw, minimum width=7mm, minimum height=0.8cm, below right=4mm of encoder] (channel) {\tiny$Channel$};
			%\node [draw, dotted, minimum size=0.8cm, right=3mm of channel] (noise) {\tiny$Noise$};
			% Receiver Section
			\node [draw, minimum width=7mm, minimum height=0.8cm, below left=4mm of channel] (decoder) {\tiny$Radio$\\\tiny$Protocol$\\\tiny$Stack$};
			\node [draw, minimum width=7mm, minimum height=0.8cm, left=3mm of decoder] (semantic2) {\tiny$Semantic$\\\tiny$Encoder$\\\tiny$P2$};
			\node [draw, minimum width=7mm, minimum height=0.8cm, left=3mm of semantic2] (output) {\tiny$Output$\\\tiny$Image$};
			% Arrows between nodes
			\draw[-stealth] (input) -- (semantic1);
			\draw[-stealth] (semantic1) -- (encoder);
			\draw[-stealth] (encoder.east) -| (channel.north);
			\draw[-stealth] (channel.south) |- (decoder.east);
			\draw[-stealth] (decoder) -- (semantic2);
			\draw[-stealth] (semantic2) -- (output);
			% Noise arrow
			\draw[stealth-, dashed] (channel) -- ++ (1.5,0) node[right]{\tiny$noise$};;
			% Transmitter Group (Dashed Border)
			\node[draw, dashed, fit=(input) (semantic1) (encoder), inner sep=0.2cm, label=above:{\scriptsize Transmitter}] (transmitter) {};
			% Receiver Group (Dashed Border)
			\node[draw, dashed, fit=(decoder) (semantic2) (output), inner sep=0.2cm, label=above:{\scriptsize Receiver}] (receiver) {};
		\end{tikzpicture}
		\caption{Splitting approach for semantic encoder.}
		\label{fig1}
	\end{figure}
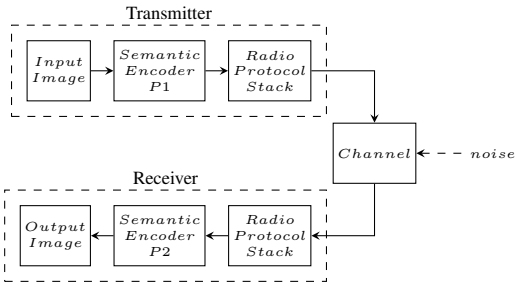

	The proposed solution provides a scalable framework for diverse communication environments that consists in splitting lightweight semantic image segmentation processing models between the transmitter, which is assumed to be a constrained resource device such as mobile, and the receiver, which is assumed to be a more powerful equipment such as an edge server. Fig.~\ref{fig1} shows the proposed method for semantic image segmentation. It has two parts, namely the "semantic encoder P1," at the transmitter side that executes part of the semantic image segmentation process, and the "semantic encoder P2," at the receiver that executes the rest of the process. Moreover, at the transmitter the channel coding together with the rest of radio protocol stack are applied to the output of the “semantic encoder P1” prior to delivering the resulting information through the channel. Similarly, at the receiver side, the received information is processed at the radio protocol stack including the channel decoding before delivering it to the “semantic encoder P2”. 
	
	\vspace{1pt}
	
	\vspace{1pt}

	Among the existing lightweight semantic image segmentation models this paper focuses on the proportional-integral-derivative network PIDNet \cite{ref09-pidnet01}. This model was chosen because it is an efficient model designed for real-time semantic image segmentation, making it well suited for this study as its balanced architecture ensures high accuracy while maintaining a low computational cost. PIDNet semantic encoder consists of six stages of semantic extraction, which allows us to split it easily.
	
	\vspace{1pt}

	Table \ref{table1} shows the stages of the PIDNet architecture. PIDNet processes the image in several steps, each one improving the representation of features while decreasing the image size. Assuming an input image with 1024 x 1024-pixel size, stage 0 applies convolutional layers to extract initial features, downscaling the image to 256 x 256 pixels and preparing the input for deeper feature extraction in subsequent stages.
	Stage 1 applies residual blocks (RB), maintaining the resolution at 256 × 256 while enhancing feature extraction. It uses skip connections to keep important features and stop gradients vanishing. Additional RBs are used in stage 2, reducing the resolution to 128 × 128 pixels and enabling more abstract feature learning while maintaining essential spatial information.
	Stage 3 introduces a three-branch structure to process features (patterns, textures, shapes, and object details extracted from the image) in parallel. The first branch extracts local features, the second downsamples and upsamples to capture contextual information, and the third refines the representations. This stage reduces the resolution to 64 × 64 and merges multi-scale features for richer information. The three-branch structure continues in stage 4 processing features more deeply while reducing the resolution to 32 × 32. The combination of extracted local and contextual information strengthens the representation of complex patterns in the image. Stage 5 employs residual bottleneck blocks (RBB), smaller version of residual blocks that make computations faster by both shrinking and expanding the sizes of features, reducing computational costs while preserving important details with the continuation of the three-branched structure. The output of this stage is a 16 × 16-pixel feature map, maintaining segmentation-relevant information at a coarse scale.
	Last but not least, stage 6 adds a PPM to capture multi-scale contextual information and creates the final 128 × 128-pixel feature map, which is then run through convolutional layers to produce the final segmentation output. 
	
	\vspace{1pt}

	\begin{table}[htbp]
		\caption{Stages of the PIDNet for semantic segmentation.\cite{ref09-pidnet01}}
		\begin{center}
			\begin{tabular}{|c c c c c|} 
				\hline
				\textbf{Stage}&\multicolumn{3}{ c }{\textbf{Operation}}&\textbf{Output} \\
				\hline
				0 & \multicolumn{3}{c}{ $2 \times Conv$ layers} & $256 \times 256$ \\
				\hline
				1 & \multicolumn{3}{c}{$RB$} & $256 \times 256$ \\
				\hline
				2 & \multicolumn{3}{c}{ $RB$} & $128 \times 128$ \\
				\hline
				3 & $RB$ & $RB$ & $RB$ & $64 \times 64$ \\
				\hline
				4 & $RB$ & $RB$ & $RB$  & $32 \times 32$ \\
				\hline
				5 & $RBB$ & $RBB$ & $RBB$ & $16 \times 16$ \\
				\hline
				\multirow{2}{*}{6} & \multicolumn{3}{c}{$PMM$} & \multirow{2}{*}{$128 \times 128$} \\
				{}&\multicolumn{3}{ c }{ $2 \times Conv$ layers}&{} \\
				\hline
			\end{tabular}
			\label{table1}
		\end{center}
	\end{table}
	
	\vspace{1pt}
	
	According to the above, we split the PIDNet after stage 5 to make sure the encoder creates small, low-resolution semantic features with rich contextual information. This splitting reduces transmission overhead while preserving task-relevant semantics for accurate reconstruction at the receiver.The splitting process after stage 5 is also easier compared to the splitting at previous stages, in which the presence of three branches complicates the separation process. Unlike typical model-splitting methods that focus on saving computing power or memory, our split is based on a communication-specific balance: reducing the bit rate while keeping important information intact, even when the channel conditions change.
	
	\vspace{1pt}
	
	Consequently, we can summarize the advantages and expected results of the splitting as follows:
	\begin{itemize}
		\item Reducing the bit rate: By transmitting images with a smaller size as shown in Table \ref{table1}. The size of the transmitted image will be 16x16 instead of 128x128. This way, we can significantly reduce the amount of data transmitted over the network, which leads to saving bandwidth.
		
		\item Reducing GPU Computing \& Memory Usage: By moving the highly computational demanding PPM process of the 6th stage and the additional two convolutional layers to the receiver, the computing demands for the resource constrained transmitter can be substantially reduced. Hence, this approach can be used on low-performance devices to perform the transmission process.
	\end{itemize}
	
	\vspace{1pt}
	
	\section{Experiments and Results}\label{expres}
	
	To evaluate the performance of our proposed solution, we conducted a series of experiments using a specially designed simulator that we programmed for this purpose. This section shows the experimental setup, describes the methodology, and presents the results, highlighting the advantages of our approach.
	
	\vspace{1pt}

	The experiments were done with a Python-based simulator. The images that are segmented have been obtained from the Cityscape dataset, which contains a variety of stereoscopic video sequences recorded in street scenes from 50 different cities \cite{ref19-cityscapesdatasetsemanticurban}. A total of 500 images are sent in each experiment. The radio transmission is modelled considering two possible modulation schemes, namely Quadrature phase shift keying (QPSK) and 16 quadrature amplitude modulation (16QAM). As such, the bandwidth of the transmitted signal depends on the amount of information to send based on the segmentation process and on the used modulation. An Additive White Gaussian noise (AWGN) channel is assumed in the link between transmitter and receiver. The effect of the AWGN channel is modelled by means of the signal-to noise ratio (SNR).  
	
	\vspace{1pt}

	The proposed solution is benchmarked against two reference schemes. Thus,  three experiments have been executed. The first one consists in the traditional communications scheme, where we send an image without any processing through the channel and then apply the complete lightweight semantic image segmentation process at the receiver. The second experiment   consists in executing the full semantic segmentation at the transmitter and the result is sent through the channel. The third experiment represents the proposed solution that splits the lightweight semantic segmentation model between the sender and the receiver as we explained in Section \ref{probsul}. In all experiments, we send the data twice, once using QPSK and the second time using 16QAM. 
	
	\vspace{1pt}

	The metrics to assess the performance of the experiments are carefully selected to capture critical aspects of semantic image segmentation, communication resources, and computational resources. They are the following ones:
	
	\vspace{1pt}

	\begin{itemize}
		
		\item The mean of Intersection over Union (mIoU): which is the average of the IoU across all categories, i.e. different object classes or labels in the dataset (e.g., road, car, pedestrian, building, sky). The IoU provides a measure of image segmentation accuracy by comparing the predicted and ground truth regions (i.e. the actual, manually labeled segmentation categories) \cite{ref20-rezatofighi2019generalizedintersectionunionmetric} and calculated as and is given:
		
		\begin{equation}
			IoU= \frac{TP}{TP+FP+FN} \label{eq1}
		\end{equation}
		Where:
		\begin{itemize}
			\item  TP is the true positive results. Refers to pixels that are correctly predicted as belonging to a specific category.
			\item  FP is the false positive results. Refers to pixels that are incorrectly predicted as belonging to a category when they do not.
			\item FN is the false negative results. Refers to pixels that belong to a category but are either unclassified or misclassified by the model.
		\end{itemize}
		
		\item Required bit rate in (Mbps): This measures the average bit rate required to be sent through the channel depending on the amount of information sent as a result of the semantic segmentation processes executed at the transmitter. 
		
		\item GPU Processing Usage: This is the percentage of GPU processing power used by the operations executed at the transmitter. 
		
		\item GPU Memory Usage: Calculates the percentage of GPU memory (VRAM) used by the operations in the transmitter for tasks such as storing models, results, or input data.
		
	\end{itemize}
	
	\vspace{1pt}

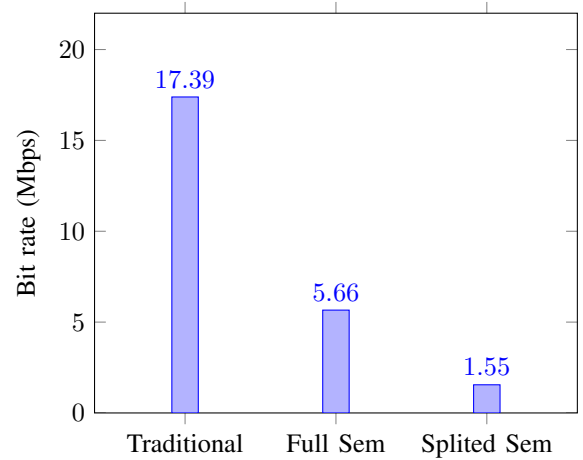
\begin{figure}[htbp] % Use [ht] for preferred placement
		\centering
		\begin{tikzpicture}
			\begin{axis}[
				ybar,
				width=8.0cm,
				ylabel={Bit rate (Mbps)},
				symbolic x coords={Traditional, Full Sem, Splited Sem},
				xtick=data,
				ymin=0, ymax=22,
				%xticklabel style={rotate=30, anchor=east}, % Rotate x-axis labels
				nodes near coords,
				nodes near coords align={vertical},
				enlarge x limits=0.3,       % Space out the bars
				]
				\addplot coordinates {(Traditional,17.39) (Full Sem,5.66) (Splited Sem,1.55)};
			\end{axis}
		\end{tikzpicture}
		\caption{Bit rate in Mbps for different experiments.}
		\label{fig2}
		\vspace{1em} % Add extra vertical space
	\end{figure}

	Fig.~\ref{fig2} shows the bit rate required to transmit data in the three experiments. It is observed that the bit rate with the proposed approach is reduced by 91\% compared to traditional communication and by 72.6\% for the full semantic segmentation before transmission.
	
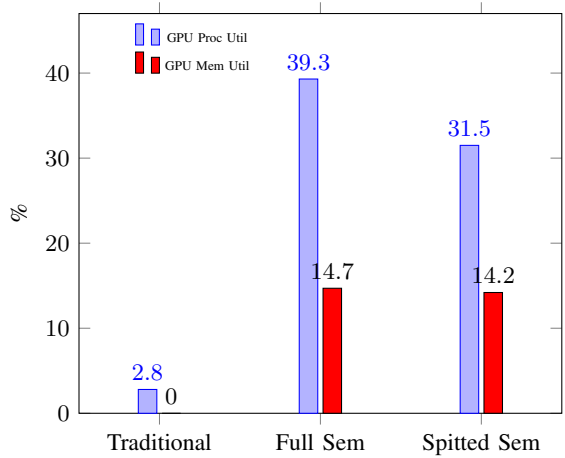
\begin{figure}[H]
		\centering
		\begin{tikzpicture}
			\begin{axis}[
				ybar,
				width=8.0cm,
				bar width=7pt,
				ylabel={\%},
				symbolic x coords={Traditional, Full Sem, Spitted Sem},
				xtick=data,
				ymin=0, ymax=47,
				%xticklabel style={rotate=30, anchor=east}, % Rotate x-axis labels
				legend style={at={(0.24,0.99)}, anchor=north, legend columns=1, font=\tiny, fill=none, draw=none}, % Position legend below chart
				nodes near coords,
				nodes near coords align={vertical},
				enlarge x limits=0.25,       % Space out the bars
				font=\small,
				]
				% GPU Comuting data
				\addplot coordinates {(Traditional,2.8) (Full Sem,39.3) (Spitted Sem,31.5)};
				\addlegendentry{GPU Proc Util}
				
				% GPU Memory data
				\addplot[fill=red] coordinates {(Traditional,0) (Full Sem,14.7) (Spitted Sem,14.2)};
				\addlegendentry{GPU Mem Util}
			\end{axis}
		\end{tikzpicture}
		\caption{GPU processing utilization and GPU memory utilization usage in percentage.}
		\label{fig3}
		\vspace{1em} % Add extra vertical space
	\end{figure}

	Fig.~\ref{fig3} shows the GPU processing utilization and GPU memory utilization in percentage. As shown, we were also able to cut down on the use of computing resources. For the full semantic segmentation before transmission, in the transmitter, the GPU processing usage dropped by 19.8\% and the GPU memory utilization dropped by 3.4\%.
	
\begin{figure}[htbp] % Use [ht] for preferred placement
		\centering
		\begin{subfigure}[htbp]{0.24\textwidth}
			\begin{tikzpicture}
				\begin{axis}[
					width=4.8cm,
					xlabel={SNR dB},
					ylabel={mIoU \%},
					legend style={
						at={(0.3,-0.4)}, % Position the legend inside the figure
						anchor=north,    % Align the legend anchor to the top-center
						draw=none,       % Remove legend box border
						fill=white,      % Add a white background to the legend
						font=\small      % Adjust legend font size
					},
					grid=both,
					]
					% Plot data1.csv
					\addplot[red, thick, mark=square] table[x=snr, y=miou_f, col sep=comma] {pidnet_qpsk_miou.csv};
					\addlegendentry{Full Sem}
					
					% Plot data2.csv
					\addplot[blue, thick, mark=o] table[x=snr, y=miou_n, col sep=comma] {pidnet_qpsk_miou.csv};
					\addlegendentry{Traditional}
					
					% Plot data3.csv
					\addplot[black, thick, mark=triangle*] table[x=snr, y=miou_s, col sep=comma] {pidnet_qpsk_miou.csv};
					\addlegendentry{Spitted Sem}
				\end{axis}
			\end{tikzpicture}
			\caption{QPSK}
		\end{subfigure}
		\hfill
		\begin{subfigure}[htbp]{0.24\textwidth}
			\begin{tikzpicture}
				\begin{axis}[
					width=4.7cm,
					xlabel={SNR dB},
					ylabel={mIoU \%},
					legend style={
						at={(0.3,-0.4)}, % Position the legend inside the figure
						anchor=north,    % Align the legend anchor to the top-center
						draw=none,       % Remove legend box border
						fill=white,      % Add a white background to the legend
						font=\small      % Adjust legend font size
					},
					grid=both,
					]
					% Plot data1.csv
					\addplot[red, thick, mark=square] table[x=snr, y=miou_f, col sep=comma] {pidnet_16QAM_miou.csv};
					\addlegendentry{Full Sem}
					
					% Plot data2.csv
					\addplot[blue, thick, mark=o] table[x=snr, y=miou_n, col sep=comma] {pidnet_16QAM_miou.csv};
					\addlegendentry{Traditional}
					
					% Plot data3.csv
					\addplot[black, thick, mark=triangle*] table[x=snr, y=miou_s, col sep=comma] {pidnet_16QAM_miou.csv};
					\addlegendentry{Spitted Sem}
				\end{axis}
			\end{tikzpicture}
			\caption{16QAM}
		\end{subfigure}
		\caption{SNR vs mIoU for (a) QPSK and (b) 16QAM.}
		\label{fig4}
		\vspace{1em} % Add extra vertical space
	\end{figure}
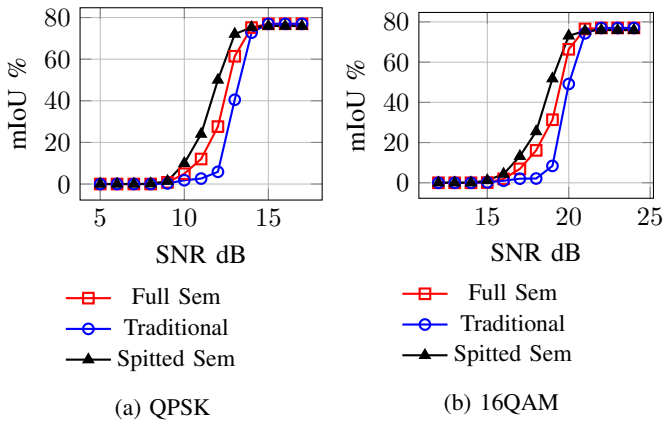

	Fig.~\ref{fig4} shows the mIoU for different executions with the two modulation schemes and for SNR values varying in the range 5-30 dB. This lets us see how the splitting process affected the accuracy of semantic segmentation. The proposed approach is able to achieve the same mIoU than the full semantic segmentation at the transmitter with 1-1.5 dB less SNR. We anticipated this outcome because of the decrease in the data rate. We also observe that under good SNR conditions the accuracy is similar to the rest of the experiments.
	
	\section{Conclusion}\label{conc}
	
	This paper has proposed the use of splitting in lightweight semantic image segmentation. This enables the model to operate on resource-constrained devices like mobiles, drones, etc. The approach intends to find a balance between communication resources and computation resources while maintaining the accuracy of image segmentation. The proposed approach has been experimentally evaluated to highlight its effectiveness. The results indicate that the required bit rate decreases by over 90\% compared to executing image segmentation entirely at the receiver and by 70\% compared to executing it entirely at the transmitter. Furthermore, the proposed approach managed to reduce the computation required for semantic image segmentation without compromising accuracy. Specifically, it reduced the GPU processing usage by more than 19\%. The promising results obtained reflect the potential of our solution to maintain segmentation accuracy while distributing the computing load across multiple devices. Although the results are promising, the simulator does not fully take into account all channel impairments, including dispersion and fast fading. Future work will focus on expanding the experiments to include the possibility of making the model lighter and more compact and applying it to 6G communication scenarios.
	
	\section*{Acknowledgment}
	The work of E. Abu-Helalah, J. Serra has been funded and supported by the ”Ministerio de Asuntos Económicos y Transformación Digital” and the European Union-NextGenerationEU in the frameworks of the ”Plan de Recuperación, Transformación y Resiliencia” and of the ”Mecanismo de Recuperación y Resiliencia” under references TSI-063000-2021-18/24/77. The work of J. Pérez-Romero has been supported by the TRAINER-6G project (PID2023-146748OB-I00) funded by MCIN/AEI/10.13039/501100011033 and by ERDF/EU.
	
	\bibliographystyle{IEEEtran}
	\bibliography{references}
	
\end{document}